\documentstyle[11pt,aaspp4]{article}

\righthead{Substellar Companion to HR 7329}
\lefthead{Lowrance et al.}

\newcommand\mj{M$_{Jup} $}

\begin{document}
 

\title{A Candidate Substellar Companion to HR 7329}

\author{Patrick J. Lowrance\altaffilmark{1}, 
Glenn Schneider\altaffilmark{2}, J. Davy Kirkpatrick\altaffilmark{3}, 
E. E. Becklin\altaffilmark{1}, Alycia J. Weinberger\altaffilmark{1}, B. Zuckerman\altaffilmark{1},
Phil Plait\altaffilmark{4}, Eliot Malmuth\altaffilmark{5}, 
Sally Heap\altaffilmark{6}, A. Schultz\altaffilmark{7}, 
Bradford A. Smith\altaffilmark{8}, 
Richard J. Terrile\altaffilmark{9}, Dean C. Hines\altaffilmark{2}}

\altaffiltext{1}{University of California Los Angeles, Los Angeles, CA}

\altaffiltext{2}{University of Arizona, Tucson, AZ}

\altaffiltext{3}{Infrared Processing and Analysis Center, California
Institute of Technology, Pasadena, CA}

\altaffiltext{4}{Advanced Computer Concepts, Inc, Goddard Space Flight Center, Greenbelt, MD}

\altaffiltext{5}{Raytheon ITSS, Goddard Space Flight Center, Greenbelt, MD}

\altaffiltext{6}{NASA, Goddard Space Flight Center, Greenbelt, MD}

\altaffiltext{7}{Computer Sciences Corporation, Baltimore, MD}

\altaffiltext{8}{Institute for Astronomy (IFA), University of Hawaii, Honolulu, HI}

\altaffiltext{9}{Jet Propulsion Laboratory (JPL), Pasadena, CA}

\begin{abstract}

We present the discovery of a candidate substellar companion from a survey of 
nearby, young stars with the NICMOS coronagraph on the Hubble 
Space Telescope. The H$\approx$12 mag object 
was discovered approximately 4$\arcsec$ from the young A0V star HR 7329.
Using follow-up spectroscopy from STIS, we derive a spectral type between M7V
and M8V with an effective temperature of $\sim$ 2600 K.
We estimate that the probability of 
a chance alignment with a foreground dwarf star 
of this nature is $\sim 10^{-8}$ and therefore
suggest the object (HR 7329B) is physically associated with HR 7329
with a projected separation of 200 AU. 
Current brown dwarf cooling models indicate
a mass of less than 50 Jupiter masses for HR 7329B based on age estimates
of $\leq$ 30 Myr for HR7329A.

\end{abstract}


\section{Introduction}

The discovery of substellar objects in stellar systems is a key goal in
contemporary astronomy, and an essential element in furthering our 
knowledge of the mass function of binary star and planetary system 
formation. The substellar mass range from 10 to 80 Jupiter
masses (0.01 $-$ 0.08 M$_{\odot}$) is crucial to our
understanding of the bridge between the lowest mass stars and 
the giant planets. 
To this end, the Near Infrared Camera and Multi-Object 
Spectrometer (NICMOS) Instrument Definition Team (IDT)  
has conducted an infrared coronagraphic 
survey of young, main-sequence stars to search for substellar 
companions. Substellar objects cool with
age because they do not sustain hydrogen fusion and
are more difficult to detect with time as they become 
fainter (e.g. Burrows et al. 1997). Using independently determined ages 
and distances for the target stars, the masses of newly-detected secondaries 
can be ascertained from infrared fluxes and 
theoretical evolutionary tracks on the H$-$R diagram. 
Follow-up spectroscopy further constrains the effective temperature
and probability of companionship. Here we present a spectrum obtained 
with the Space Telescope Imaging Spectrograph (STIS) of a substellar 
companion candidate, HR 7329B, from our NICMOS imaging survey.
Previously, this survey revealed TWA5B, a $\sim$ 20 Jupiter mass brown
dwarf companion to TWA 5A (Lowrance et al. 1999).

\section{NICMOS}

\subsection{Observations}

HR 7329 (HD 181296; A0V; d$\sim$48pc; V=5.05; RA=19h 22m 51.2s, DEC=$-$54$^{\circ}$ 25$'$ 26$''$ (J2000.0)) 
was observed with NICMOS on 1998 June 29, 16:15 $-$ 17:17 UT. 
We obtained multiple-exposure images with the star
behind the coronagraph (radius = 0.3$\arcsec$) on 
Camera 2 (pixel scale  = $\sim$0.076 arcsec pixel$^{-1}$) and a 
wide-band F160W 
filter (central wavelength: 1.5940~$\mu$m, $\Delta\lambda$ = 0.4030~$\mu$m), which 
corresponds closely to the central wavelength of a Johnson H-band photometric filter. 
Five standard NICMOS STEP16 MultiAccum (non-destructive read) 
integrations (MacKenty et al. 1997) totaling 719.6s were executed at
each of two orientations differing
by 29.9 degrees. While the stellar point-spread-function (PSF), 
the instrumental scattering function, and detector artifacts 
rotate with the aperture, any real features in the unocculted area of the 
detector will be unaffected by a change in the camera orientation. Subtraction 
of these two images has been shown to significantly reduce residual PSF background
light (Schneider et al. 1998). 
The NICMOS coronagraphic images were reduced and processed
utilizing calibration darks and flat-fields created by the 
NICMOS IDT from on-orbit 
observations following the method described in Lowrance et al. (1999).

\subsection{Results}

Subtraction and analysis of the NICMOS coronagraphic images reveal a 
stellar-like object (HR 7329B) at 
a separation of 4.17$\arcsec$ $\pm$ 0.05, and a position angle of 
166.8$^{\circ}$ $\pm$ 0.2 
from HR 7329 (HR 7329A) (Figure 1). This secondary is point-like with
a Full Width Half Maximum (FWHM) of 0.$\arcsec$15 (the
diffraction limit is 0.$\arcsec$14) with the first Airy ring apparent in
Figure 1.  Since the target star is occulted in 
the NICMOS coronagraphic images, its position is ascertained from the target 
acquisition image and located behind the coronagraph by a known
telescope offset. 

\placefigure{fig1}

The secondary fell near the edge of the field of view in the second
orientation, so the magnitude of HR 7329B was measured 
using a 12 pixel radius circular aperture 
centered on the companion in the subtracted image from 
the first orientation.  A correction
factor of 9.66$\%$, determined 
from coronagraphic photometric aperture corrections
developed by the NICMOS IDT, was applied to compensate for
the flux which fell out of this aperture. The [F160W] magnitude of HR 7329B
is then 11.90 $\pm$ 0.06 mag using a conversion 
factor for the F160W filter of 2.19 $\times\ 10^{-6}$ Jy ADU$^{-1}$s, and 
1083 Jy corresponding to an H magnitude of zero in the Vega system
(Rieke 1999) where the majority of the uncertainty is dominated in
NICMOS's calibration in relation to standard stars. 
The F160W filter is $\sim 30\%$ wider than ground-based Johnson H-band
filters which necessitates a careful conversion from F160W to H band for
cool temperature objects. For six M 
dwarfs between spectral types M6 and M9 with measured
F160W and ground-based H-band magnitudes, we find a mean difference
of 0.03 $\pm$ 0.02 mag. For HR 7329B, a M7.5, we thus expect the
H-[F160W] color to be about 0.03 mag. Making this color correction, we
estimate an H magnitude of 11.93 $\pm$ 0.06 mag.

The [F187N] magnitude of HR 7329A was determined from aperture 
photometry of the two calibrated target acquisition images (at each 
of the two spacecraft orientations) processed as described in Lowrance
et al. (1999). Within the uncertainties, the two measurements agreed and 
were averaged to yield [F187N] $=$ 5.0 $\pm$ 0.1 mag.

\section{STIS}

\subsection{Observations}

HR 7329 was acquired into the STIS 52$\arcsec\times$0.$\arcsec$2 slit
on 20 May 1999 
and then offset by $0.95\arcsec$ in right ascension and $-4.06\arcsec$ in declination
(based on the NICMOS astrometric results) 
to place the secondary into the slit. To keep the
primary as far out of the slit as possible, we employed a slit
position angle of 252.06 degrees so that the line joining 
the primary and secondary was approximately perpendicular to the slit, 
thereby minimizing contamination from scattered primary light.  
Spectral imaging sequences were
completed in one orbit with the G750M grating in three tilt settings
with central wavelengths of 8311,
8825 and 9336\AA\ (resolution $\sim$ 0.55\AA) 
for total integration times of 340, 172, and 150 seconds, respectively. 
At each tilt setting we executed a two position dither of 0.$\arcsec$35 along the
slit to allow replacement of bad or hot pixels, and the 
exposures were split for cosmic ray removal. Thus, we obtained four spectra
at each tilt setting. After each set of four spectral images, we obtained
flat fields required to calibrate the known effects of fringing which 
appear longward of $\sim$ 7500\AA\ . Due to a failure of HST to acquire
one of the two guide stars, there was a small differential pointing error of
about 0.04\arcsec, or 1 pixel. This caused the secondary to be
marginally de-centered and as a result a small percentage of the target flux
fell out of the slit.

\placefigure{fig2}

\subsection{Results}

The STIS spectra were calibrated, averaged, binned to a resolution of
$\sim$ 6\AA\ and normalized to the flux (in ergs/s/cm$^2$/\AA) at 8500\AA. 
We compared the final, total spectrum to those of standard low-temperature dwarf and 
giant star spectra with a resolution = 18 \AA, a factor of three lower
than our STIS spectrum
(Kirkpatrick et al. 1991; Kirkpatrick et al. 1997) (see Figure 2). 
The HR 7329B spectrum contains an absorption line near
8200\AA\ which we attribute to the Na I doublet, which does not appear in
late-type giant stars, but is nicely fit in the dwarfs. Also, 
as seen in Figure 2, the slope of the
spectrum from 8600 to 8850\AA\ is small, as in the dwarf spectra, whereas
it rises sharply for giant stars. The NaI line is fit very well by the
M8 V standard, but the TiO absorption near 8860\AA\ is best fit by the
M7 V spectrum. We therefore assign HR 7329B a spectral type of M7.5 V with
an uncertainty of 0.5 spectral type.    
 
The diffraction spikes from the primary star also fall in the slit above
and below the secondary, and we used the relative positions of the three
resulting spectra to determine the primary-secondary separation.  The
result of 4.13 $\arcsec$ $\pm$ 0.05 agrees, to within the
uncertainties, with the NICMOS measured separation reported in Section 2.2.

\section{Discussion}

\subsection{Likelihood of Companionship}

From its H-magnitude and M7.5V spectral type, HR 7329B 
can be either a background object, a foreground main-sequence M star, 
or a companion to HR 7329A. A main-sequence M7.5V star has 
M$_{H}$ $=$ 10.3 (Kirkpatrick \& McCarthy 1994), so HR 7329B is too
bright to be a background main-sequence star. If it were on the
main-sequence, its photometric distance would be 19 parsecs. Henry (1991), 
in a volume limited infrared survey, finds six objects with M$_{H}$ $>$ 9.5 within 
five pc from the sun. If we assume a 
spherical distribution of low mass stars in the solar neighborhood, 
we can extrapolate the results within 5 pc to expect 
1000 such objects out at 25 pc, so the a priori probability
of finding one in projection within a 4$\arcsec$ radius circle is $\sim 10^{-7}$. 

Proper motion measurements of the companion and primary in the time
between the NICMOS and STIS observations could be used to further
constrain the probability of companionship. Unfortunately, the
positional errors are too large.
However, we can further constrain the probability that the object is
not a foreground M dwarf. Searching the Tycho cataloque,
we find that the mean proper motion of 1000 stars between 16 and 25
parsecs is 0.373$\arcsec \pm$ 0.277. Therefore, if we assume a
gaussian distribution of proper motions about this mean, almost 80\%
of foreground stars have moved more than the half-width of the STIS slit 
(0.1$\arcsec$)(taking into account angles along the slit), 
and would not be visible in the second epoch.

Given these arguments, it is unlikely ($\sim 10^{-8}$) HR 7329B is a foreground
object and for the remainder of the paper, we assume it is
physically associated with HR 7329A.

\subsection{Age of System}

It is difficult to determine an age for A-type stars, but HR 7329 appears to be
young ($<$ 40 Myr) based on its rotation, and more importantly, 
location on an H$-$R diagram. 
For massive stars, rotational velocities decline with age; 
HR 7329 has an especially large v$sini$ ($=$ 330 km/s) (Abt \& Morrel
1995) which is considerably above 
the majority of A-type stars ($\sim$ 100 km/s). 
Figure 3 reproduces the H$-$R diagram from Jura et
al (1998) for A stars from the Yale Bright Star Catalog and overplots 
nearby, young clusters. There seem
to be common areas of similar age stars; the 50-90 Myr IC2391 and
Alpha Per clusters lie below the older (600 Myr) Hyades and Preasepe. There is
a large scatter in the Pleaides (70-125 Myr), 
which could be due to a range of distances and 
ages as well as unresolved binaries.
HR 7329 lies on a line located below the Alpha Per
and IC 2391 cluster which intersects $\beta$ Pic, HR 4796 and HD 141569. 
The latter stars have recently been assigned ages from their late-type
companions of 20, 8, and 4 Myr respectively (Barrado Y Navascues et al. 1999;
Stauffer, Hartmann, \& Barrado Y Navascues 1995; 
and Weinberger et al. 2000). This suggests that HR 7329 is between 10
and 30 Myrs old. 
Finally, it has recently been suggested that HR 7329 is found within a young
co-moving cluster much like the TW Hydrae Association with an age of $\sim$40 Myr 
(Zuckerman \& Webb 2000; Webb et al. 2000). 

\placefigure{fig3}

\subsection{Effective Temperature and Bolometric Luminosity}

An effective temperature of HR 7329B is required to 
position it on an H$-$R diagram, but the temperature 
scale for late, young M-dwarfs is uncertain (Allard et al. 1997). 
Luhman \& Rieke (1998) extrapolate from Leggett et al.'s (1996) model
fits to derive 2670 K and 2505 K for M7 V and M8 V respectively, 
which agrees with the newer models used by Leggett, Allard, \&
Hauschildt (1998) with an uncertainty of about 100 K.
With this uncertainty for late M dwarf stars and the added
uncertainty due to the spectral type, we plot the derived temperatures
for each spectral class (Figure 4) and their associated uncertainty 
which overlaps and gives a possible range from 2405 K to 2770 K.

The parallactic distance measured to HR 7329A by the Hipparcos mission
is 47.67 $\pm$ 1.6 pc. With a derived H magnitude of 11.93 for HR7329B, and 
a distance modulus of 3.39, we calculate an absolute H magnitude of 8.54
mag. There exists a number of bolometric corrections in the literature
for M7 V and M8 V stars (Tinney et al. 1993; Kirkpatrick et al. 1993;
Bessel, Castelli, \& Plez 1998) based on I and K magnitudes. 
However, none give the BC in the H band. We have used the BC at the
other bandpasses and the colors of late-type stars as a function of
spectral type from Kirkpatrick \& McCarthy (1994) to find a
relationship between BC(H) and spectral type. For M7 and M8 we find a
range of BC$_{H}$ from 2.54 to 2.78.
Using a solar M$_{bol}$ of 4.75, we derive a luminosity for HR 7329B
of 0.0026 $\pm$ 0.0003 L${\odot}$, with an uncertainty that includes   
the 0.5 spectral type range, bolometric correction, and distance errors.

\subsection{Derived Mass}

We place HR 7329B on pre-main sequence 
evolutionary tracks (Baraffe et al. 1998) to 
infer a mass (Figure 4). Assuming only companionship, and therefore
distance, indicates a mass of 
less than 50 \mj\ (less than 35\mj\ is not covered in Baraffe's models) 
and an age of less than 30 Myr. This supports  
the young age attributed to HR 7329A from its position on the H$-$R 
diagram, other youth indicators and possible membership in 
a young moving group. Evolutionary tracks 
from different authors do differ somewhat due to the different model 
atmospheres used. The tracks of D$'$Antona \& Mazitelli (1997)
indicate a mass range of 40 \mj\ or less for this luminosity and temperature.
Burrows et al.'s (1997) models predict a 40 \mj\ brown dwarf 
will have an effective temperature
of 2800K and a luminosity of 0.0023 L${\odot}$ at an age of 22 Myr.

\placefigure{fig4}

\section{Limits on Disk Detection}

To look for possible reflected light from a circumstellar
disk around the primary,
we subtracted an observed coronagraphic PSF from each roll of HR 7329.
The NICMOS PSF is time variable, exhibiting small-amplitude
structural changes over multi-orbit timescales (Kulkarni et al. 1999). 
To find the
best matched coronagraphic PSF to HR 7329, we tested each observation of the
40 other stars in our NICMOS program to see which gave the
lowest noise subtraction as measured in the diffraction spikes and in
annuli from 0.3$-$4$\arcsec$.  There was no evidence of excess scattered
light from a disk in any of the subtractions, but the first visit of the
star HD 17925 observed on 26 Sep 1998 gave the lowest subtraction
residuals.  This K1~V star is 2.3 times brighter than HR 7329 at F160W.
A plot of the azimuthally averaged residual surface brightness after
subtraction is shown in the top panel of Figure 5.
The error bars represent the standard deviation of all of the pixels
(not including pixels obscured by diffraction spikes) averaged at each
radius.  The residuals are everywhere consistent with zero, i.e. no disk
detection.  The lower panel shows these uncertainties, multiplied by three
and converted to F160W magnitudes as a measure of the disk flux which
could have been detected at each radius.

\placefigure{fig5}

HR 7329 appears in the IRAS Point Source catalog as having excess
thermal infrared emission, indicating orbiting dust (Mannings \&
Barlow 1998). After color correcting the
catalog fluxes for spectral index and subtracting the stellar photospheric
contribution, the flux densities are F$_{12\mu m}$=0.25$\pm$0.09 Jy,
F$_{25\mu m}$=0.36$\pm$0.05 Jy, F$_{60\mu m}$=0.52$\pm$0.05 Jy, and an
upper limit of 1 Jy at 100$\mu$m.  These give a total dust optical depth
$\tau$=$L_{IR}/L_{star}\approx 3.5 \times 10^{-4}$, which is an
order of magnitude smaller than other similar stars at comparable
distances such as HR 4796 and HD 141569 around which NICMOS imaged
disks (Schneider et al. 1999, Weinberger et al. 1999).

\section{Discussion}

We present high signal-to-noise 
near-infrared photometry and optical spectroscopy of
a probable companion (HR 7329B) at a projected distance of 200 AU 
from HR 7329 (A). We suggest the mass
of B is less than 40 \mj\ . The derived age of
less than 30 Myr for this companion supports the very young age 
of the primary A0V star indicated by its placement on the H$-$R diagram
of nearby A-type stars. We do not detect any 1.6$\micron$ scattered 
light from the far-infrared emitting dust seen by IRAS 
around HR 7329A.

The HR 7329 system stands out from other binaries in that it has a
very high mass ratio, q$\sim$0.01. Zuckerman and Becklin (1992) found that
around $\sim$ 200 white dwarf stars whose progenitors are F and A main
sequence stars, the percentage of systems with low-mass M star
companions (M$\sim$0.1M$_{\odot}$, q$\sim$0.06) was 5 to 10\%, and the number of
detected brown dwarfs was one (GD 165B), or $<$1\%. The small
percentage of white dwarfs with detectable brown dwarf companions 
is probably the result of the decline in brown dwarf luminosity with age.
The discovery of 
the brown dwarf, HR 7329B, among a small sample of young A and F stars
($\sim$10) observed by NICMOS suggests 
that the number of companion brown dwarfs and low mass
stars may not be too different. In the field (Reid 1999), and the
Pleiades cluster (Zapatero Osorio et al 1997), the relative number of low mass
stars and brown dwarfs per log mass interval is also about equal,
suggesting a flat inital-mass-function (IMF) for single stars. 
Clearly, greater statistics are needed
before firm conclusions can be reached about the IMF of secondaries.

We would like to thank M. Jura, C. Chen and J. Patience for their invaluable help
and assistance. We thank the anonymous referee for comments which
help to clarify the presentation. This work is supported in part by
NASA grants NAG 5-4688 to UCLA and 
NAG 5-3042 to the University of Arizona NICMOS Instrument Definition Team.  
This paper is based on observations obtained under grant
GO-8176.01-97A with the NASA/ESA Hubble Space Telescope 
at the Space Telescope Science Institute, which is operated by the
Association of Universities for Research in Astronomy, Inc. under NASA contract
NAS 5-26555.

\clearpage

{\centering \leavevmode
\epsfxsize=.95\columnwidth \epsfbox{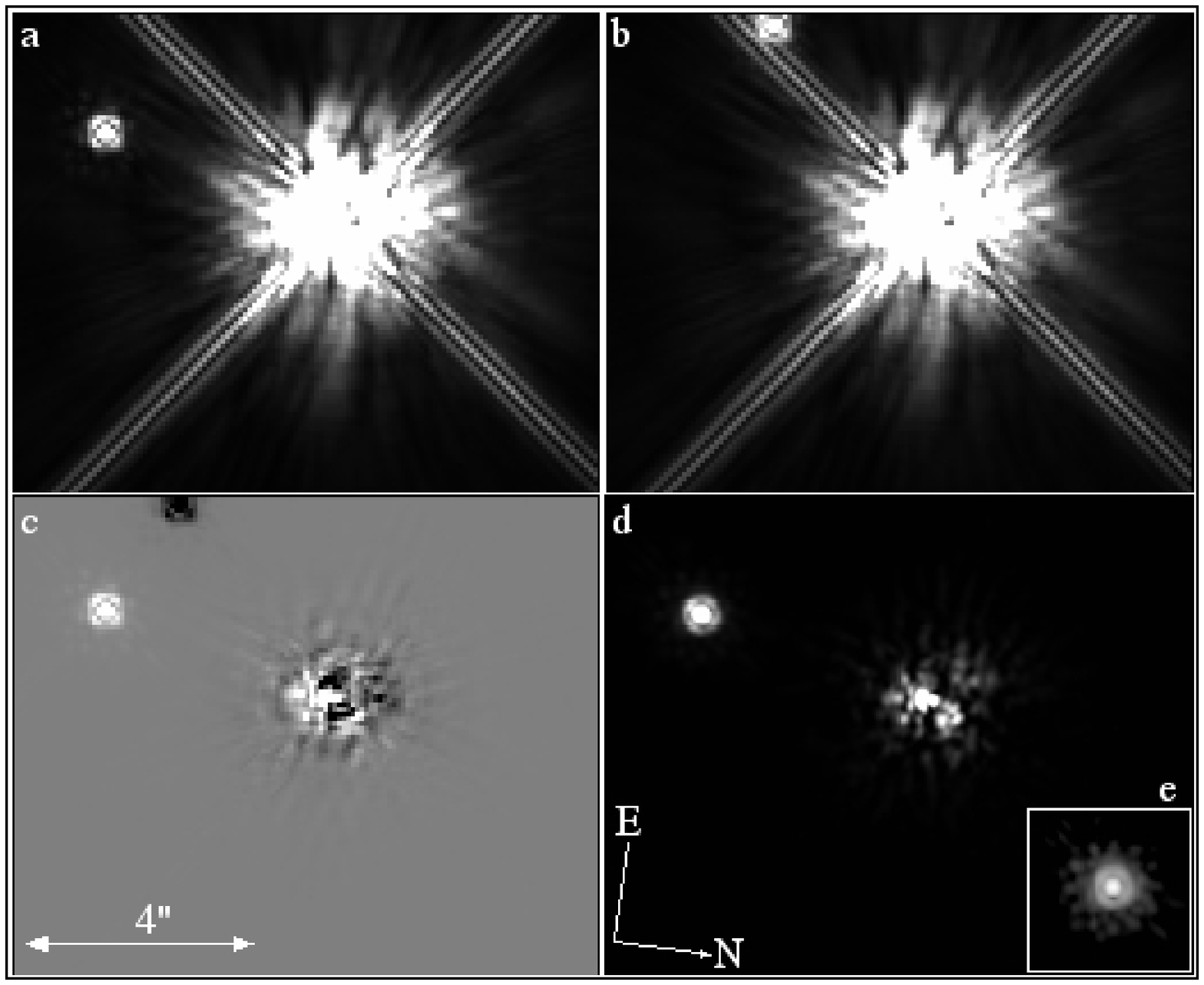}}

\figcaption[lowrance1.ps]{ The NICMOS H-band image of HR 7329B. 
Observations at two different roll orientations ({\it a} and {\it b}) 
have been subtracted with HR 7329A behind the
coronagraph ({\it c}). The diffraction spikes were masked 
and the images were rolled about the primary center and co-added ({\it d}), 
leaving a sub-sampled image ({\it e}) of HR 7329B.    \label{fig1}  }

{\centering \leavevmode
\epsfxsize=.95\columnwidth \epsfbox{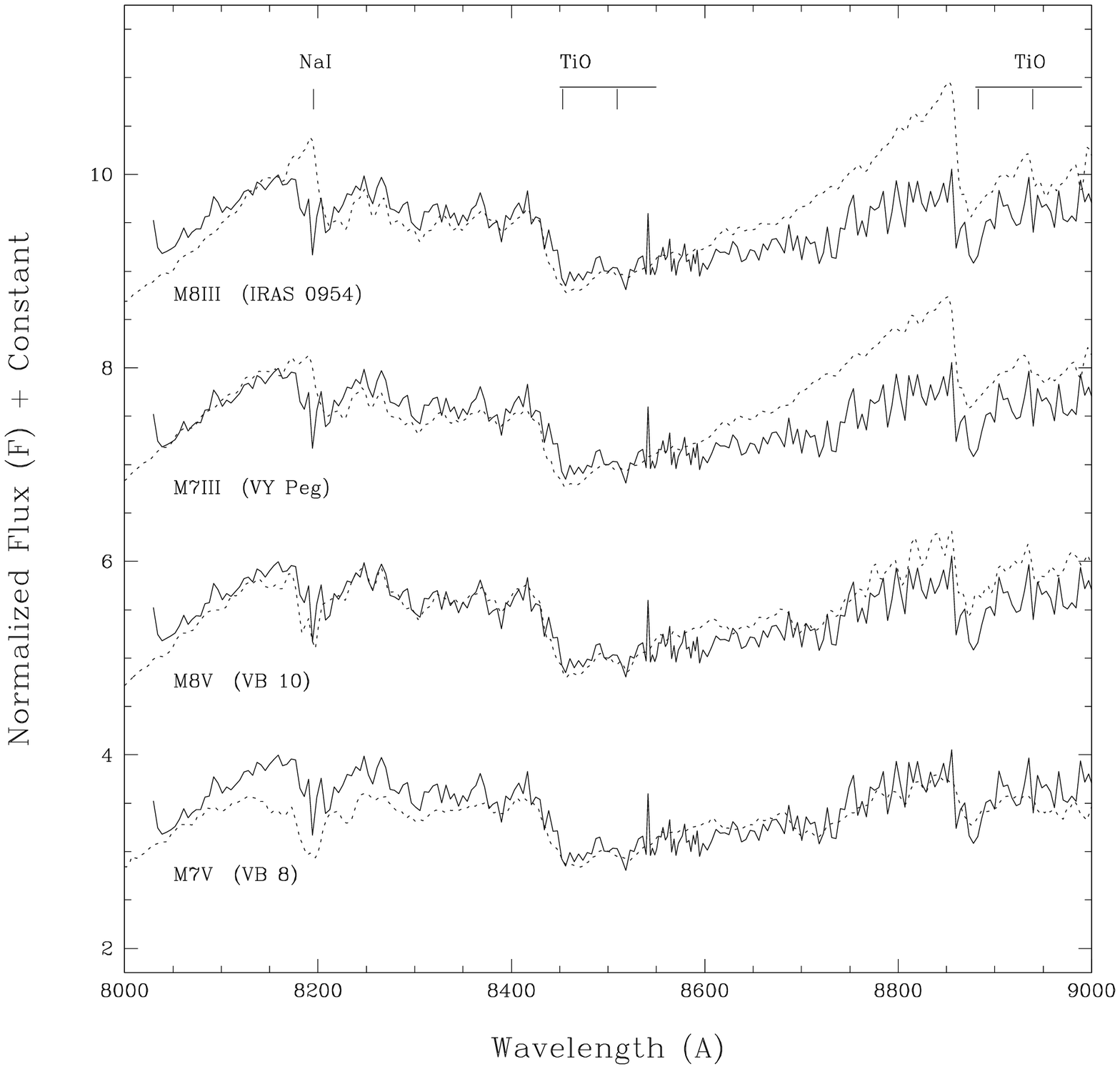}}

\figcaption[lowrance2.ps]{ STIS spectrum of HR 7329B (solid) (in
ergs/s/cm$^2$/\AA)
compared with standard late-type M dwarf and giant spectra (dashed)
(Kirkpatrick at al 1991;Kirkpatrick et al. 1997). The giant spectra
fit neither the NaI absorption near 8200\AA\ nor the
the slope later then 8600\AA . The best fit lies between
M7V and M8V. (The longward cutoff of 9000\AA\ is where the SNR of the
STIS spectrum becomes too low due to fringing effects.   \label{fig2}  }

{\centering \leavevmode
\epsfxsize=.95\columnwidth \epsfbox{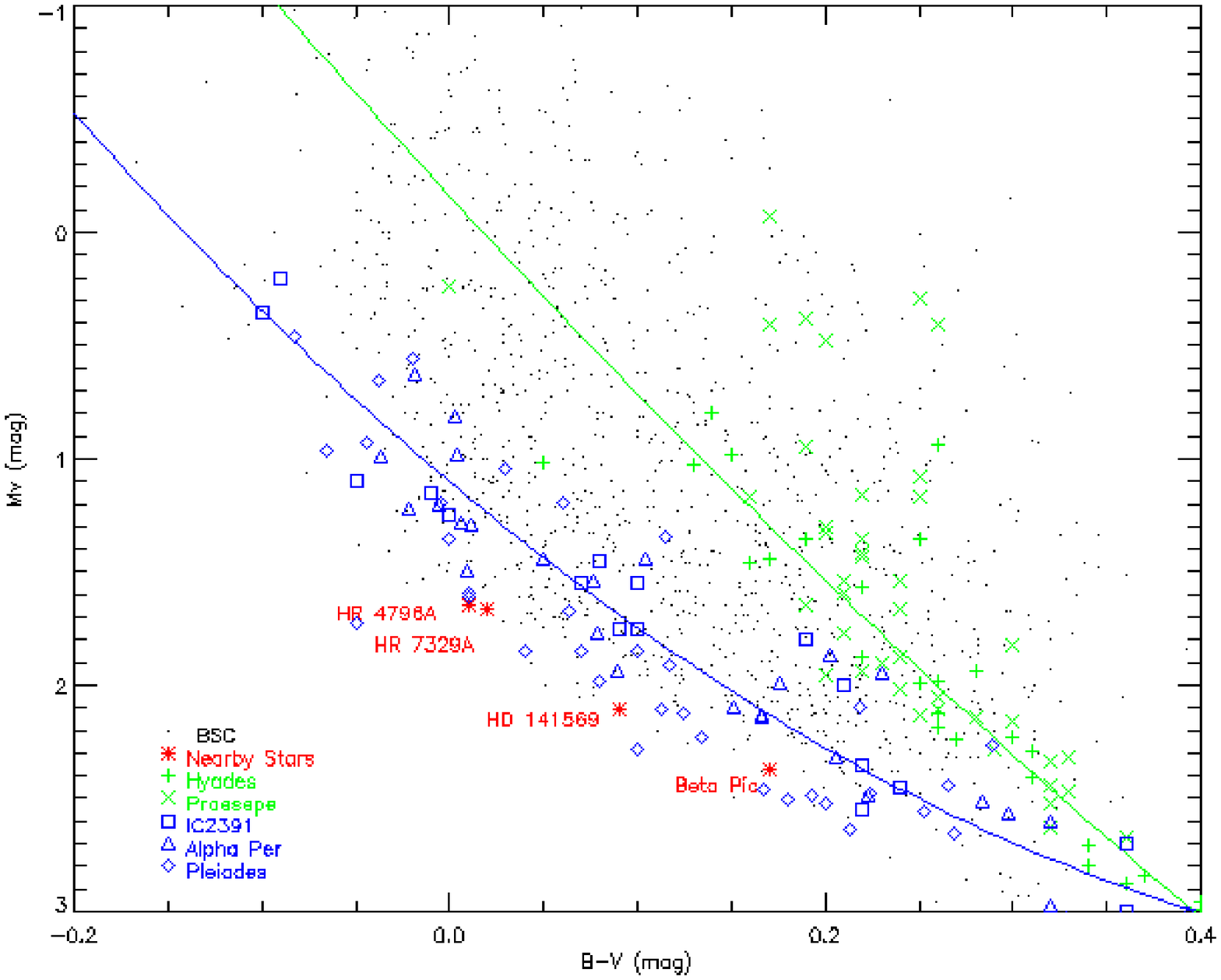}}

\figcaption[lowrance3.ps]{ H$-$R diagram for A-type stars in the Yale
Bright Star Catalogue reproduced from
Jura et al. (1998) with the nearest star clusters plotted. The lines
indicate common centers for Hyades/Preasepe (600 Myr) and Alpha Per/IC2391
(50$-$90 Myr) The asteriks indicate HR 7329 and the nearby stars assigned young ages
(4$-$20 Myr) from late-type companions.   \label{fig3}  }

{\centering \leavevmode
\epsfxsize=.95\columnwidth \epsfbox{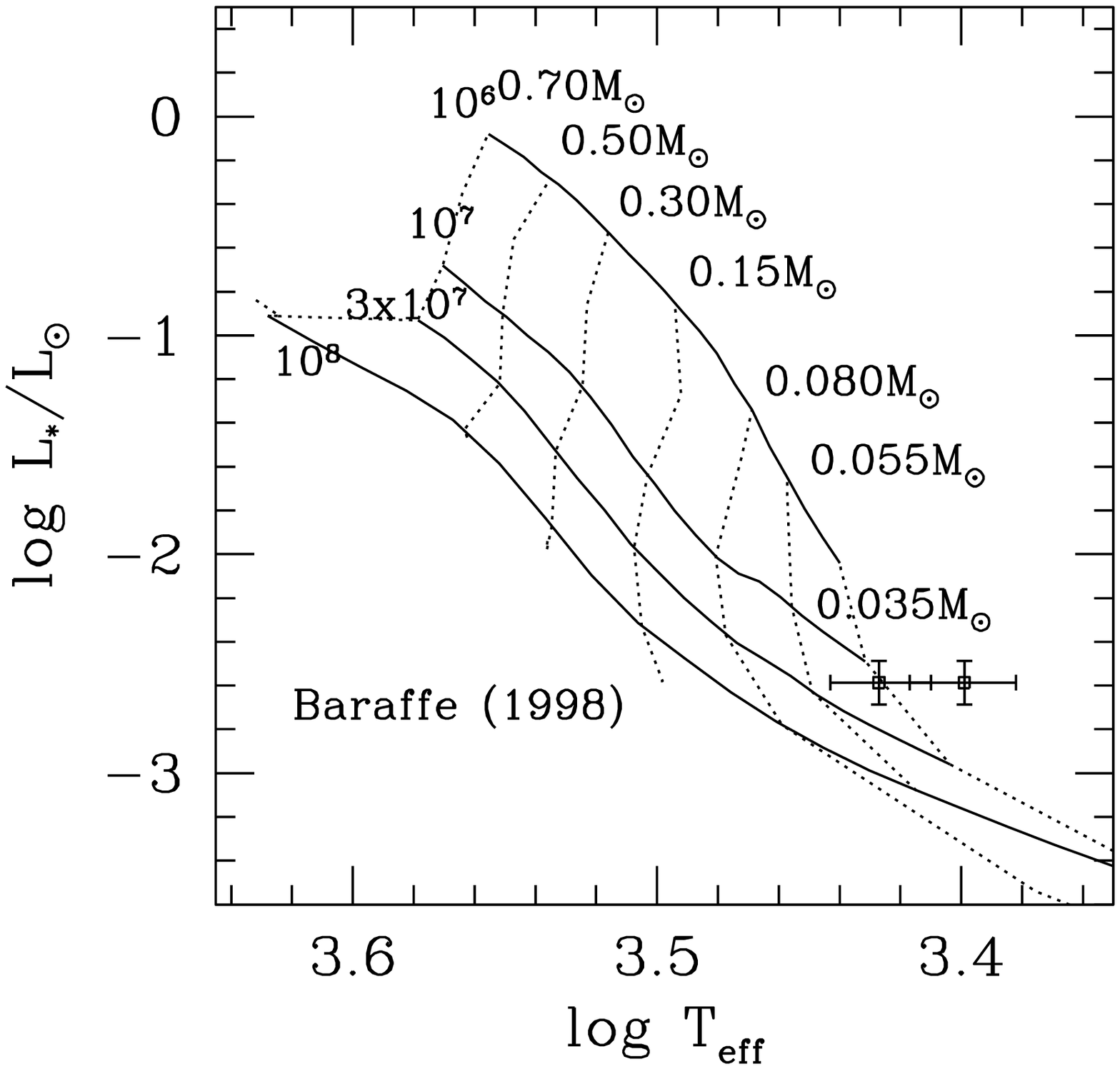}}

\figcaption[lowrance4.ps]{ Evolutionary tracks (Baraffe et al. 1998)
with HR 7329B (squares) plotted at the derived temperatures for M7 V
and M8 V with the uncertainty (which overlap) in assigning a
temperature to a low mass star (see text for discussion). 
From these and other evolutionary models we derive an age 
less than 30 Myr for this pair and a mass of less than 50 \mj\ for the secondary.   \label{fig4}  }

{\centering \leavevmode
\epsfxsize=.95\columnwidth \epsfbox{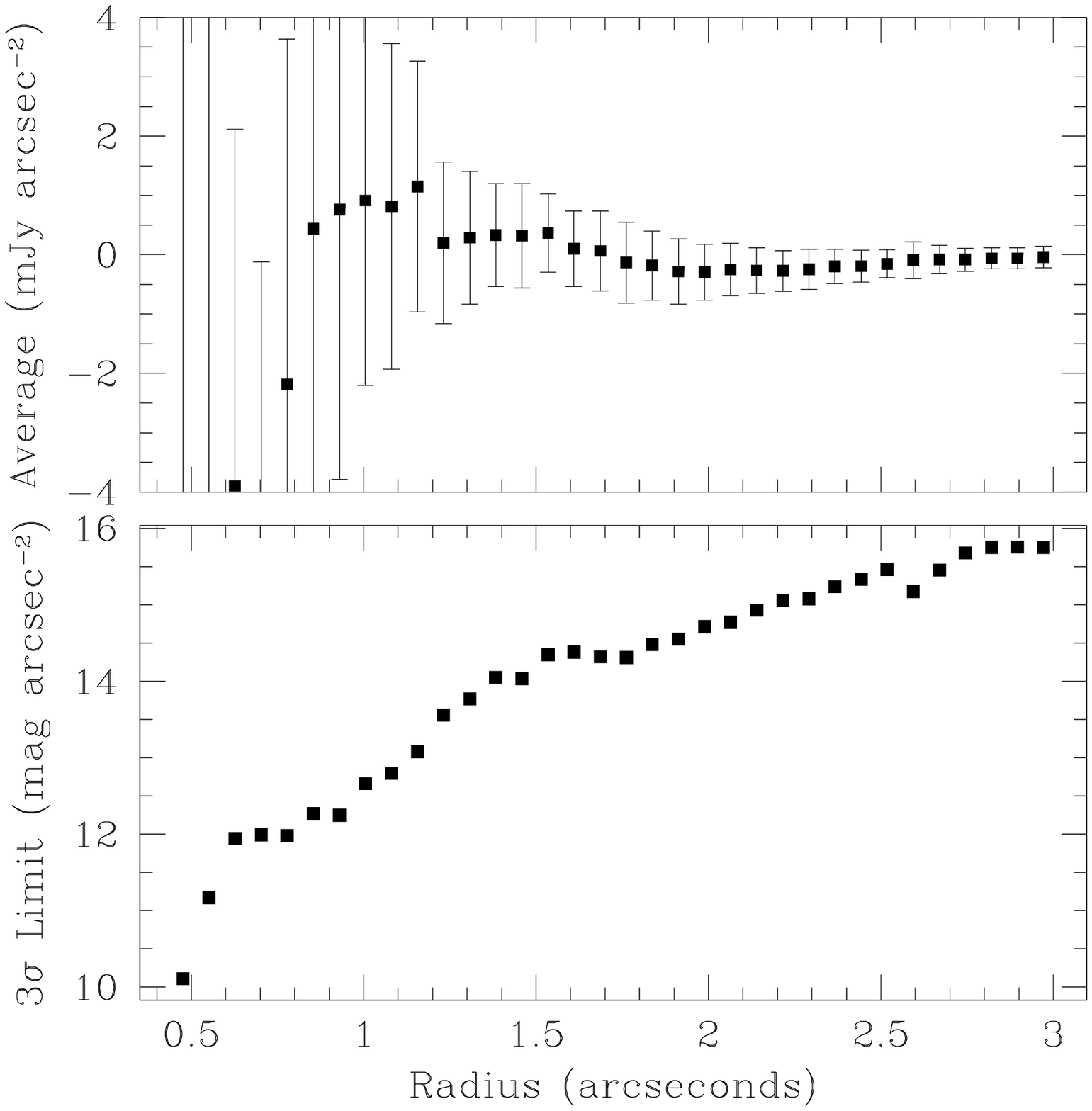}}

\figcaption[lowrance5.ps]{ Upper panel: A plot of the azimuthally
averaged residual surface brightness after
subtraction of a coronagraphic PSF. Bottom panel: 3-sigma limits on
detection of disk flux as a function of distance from HR7329A.    \label{fig5}  }

\end{document}